\documentclass[5p,times,twocolumn]{elsarticle}
\usepackage{amsmath,amssymb}
\biboptions{comma,sort&compress}
\usepackage[utf8]{inputenc}
\usepackage{graphicx}
\usepackage{bm}
\usepackage{hyperref} 
\usepackage[dvipsnames]{xcolor}
\hypersetup{
colorlinks = true,
linkcolor = blue,
anchorcolor = blue,
citecolor = blue,
filecolor = blue,
urlcolor = blue
}
\usepackage{ulem}
\graphicspath{{fig/}}
\def\p{{\cal P}}
\def\e{{\cal E}}
\def\taur{\tau_{\rm R}}
\def\taueq{\tau_{\rm eq}}

\journal{Physics Letters B}

\begin{document}

\begin{frontmatter}

\title{Extended relaxation time approximation and relativistic dissipative hydrodynamics}

\author[niser]{Dipika Dash}
\ead{dipika.dash@niser.ac.in}
\author[niser]{Samapan Bhadury\corref{cor1}}
\ead{samapan.bhadury@niser.ac.in}
\author[tifr]{Sunil Jaiswal\corref{cor1}}
\ead{sunil.jaiswal@tifr.res.in}
\author[niser]{Amaresh Jaiswal\corref{cor1}}
\ead{a.jaiswal@niser.ac.in}
\cortext[cor1]{Corresponding author}

\address[niser]{School of Physical Sciences, National Institute of Science Education and Research, An OCC of Homi Bhabha National Institute, Jatni 752050, Odisha, India}
\address[tifr]{Department of Nuclear and Atomic Physics, Tata Institute of Fundamental Research, Mumbai 400005, India}
\date{\today}

\begin{abstract} 
Development of a new framework for derivation of order-by-order hydrodynamics from Boltzmann equation is necessary as the widely used Anderson-Witting formalism leads to violation of fundamental conservation laws when the relaxation-time depends on particle energy, or in a hydrodynamic frame other than the Landau frame. We generalize an existing framework for consistent derivation of relativistic dissipative hydrodynamics from the Boltzmann equation with a energy-dependent relaxation-time by extending the Anderson-Witting relaxation-time approximation. We argue that the present framework is compatible with  conservation laws and derive first-order hydrodynamic equations in landau frame. Further, we show that the transport coefficients, such as shear and bulk viscosity as well as charge and heat diffusion currents, have corrections due to the energy dependence of relaxation-time compared to what one obtains from the Anderson-Witting approximation of the collision term. The ratio of these transport coefficients are studied using a parametrized relaxation-time, and several interesting scaling features are reported.
\end{abstract}


\end{frontmatter}


\section{Introduction}

Relativistic Boltzmann equation is a transport equation governing the space-time evolution of the single particle phase-space distribution function, and is capable of accurately describing the collective dynamics of the system. However, the Boltzmann equation is difficult to solve directly as the collision term depends on the integral of the product of distribution functions, making it a complicated integro-differential equation. Many approximation for simplifying the collision term in the linearized regime have been proposed over several decades \cite{Bhatnagar:1954zz, Welander, marle1969, anderson1974relativistic}. The Anderson and Witting relaxation time approximation (RTA) of the collision term \cite{anderson1974relativistic} enormously simplifies the Boltzmann equation by assuming that the effect of collisions are to exponentially drive the system towards local equilibrium, controlled by a single parameter--`relaxation' time.
This approximation has been used quite successfully in several fields of physics. 
It has been employed to formulate relativistic dissipative hydrodynamics and derive transport coefficients \cite{Muronga:2006zx, York:2008rr, Betz:2008me, Romatschke:2009im, Denicol:2010xn, Denicol:2012cn, Jaiswal:2013npa,  Jaiswal:2014isa, Denicol:2014loa, Gabbana:2017uvc, Mohanty:2018eja, Kurian:2020qjr}, and has also been recently used to study the domain of applicability of hydrodynamics and investigate the hydrodynamization phenomena \cite{Florkowski:2013lya, Denicol:2014xca, Denicol:2014tha, Heinz:2015gka, Heller:2016rtz, Blaizot:2017ucy, Florkowski:2017olj, Strickland:2018ayk, Kurkela:2019set, Blaizot:2020gql, Dash:2020zqx, Jaiswal:2020hvk, Soloviev:2021lhs, Chattopadhyay:2021ive, Jaiswal:2021uvv}.

The usual formulation of hydrodynamics from RTA assumes the relaxation-time to be independent of particle energy (or momentum), and one is restricted to work in Landau frame to ensure macroscopic conservation laws. However, the  collision time scale typically depends on the microscopic interactions for any realistic system \cite{Dusling:2009df, Chakraborty:2010fr, Dusling:2011fd, Teaney:2013gca, Kurkela:2017xis}. Promoting the relaxation-time to be particle energy-dependent or employing a general matching condition leads to violation of microscopic conservation laws, which has generated much interest towards a consistent approximation of the collision term which satisfies microscopic and macroscopic conservation laws \cite{Pennisi_2018, CARRISI2019298, Mitra:2020gdk, Rocha:2021zcw, Hoult:2021gnb}. Recently, a modification of the RTA was proposed \cite{Rocha:2021zcw} to ensure microscopic conservation of particle number and energy momentum for energy-dependent relaxation-time irrespective of choice of hydrodynamic frames. In the present work, we consider a different resolution to the problem \cite{Teaney:2013gca}. 
By defining the equilibrium distribution function appearing in the RTA approximation in the ``thermodynamic frame'' \cite{Banerjee:2012iz, Jensen:2012jh, Kovtun:2019hdm}, we develop a framework for consistent derivation of order-by-order hydrodynamics where conservation laws hold at each order in the gradient expansion. 
Within this framework, we derive first-order hydrodynamic equations in Landau frame with an energy-dependent relaxation-time and show that the transport coefficients, such as shear and bulk viscosity as well as charge and heat diffusion currents, have corrections due to the energy dependence of relaxation-time. 
We further study the ratios of these transport coefficients using a power law parametrization for the energy dependence of the relaxation-time and find several new and interesting scaling features that we report here.

\section{Relativistic hydrodynamics}

Conserved energy momentum tensor and net-charge current can be expressed in terms of the single particle phase–space distribution function:
\begin{align}
T^{\mu\nu} &= \!\int\! dP\, p^\mu p^\nu \!\left( f+\bar{f} \right)\! = \e u^\mu u^\nu -\left(\p+\Pi\right)\! \Delta^{\mu\nu} + \pi^{\mu\nu} \label{cons_emt} \\ 
N^\mu &= \!\int\! dP\, p^\mu \left( f-\bar{f} \right) = n u^\mu + n^\mu, \label{cons_curr}
\end{align}
where $dP = g\, d\textbf{p}/[(2\pi)^3 p_0]$ is the invariant momentum-space integration measure with $g$ being the degeneracy factor and $p^\mu$ is the particle four momenta. Here $f$ and $\bar{f}$ are the phase-space distribution functions for particles and anti-particles respectively. In the tensor decomposition, $\e$, $\p$, and $n$ are the energy density, equilibrium pressure, and the net number density. The projection operator $\Delta^{\mu\nu}=g^{\mu\nu}-u^\mu u^\nu$ is orthogonal to the hydrodynamic four-velocity $u^\mu$ defined in the Landau frame: $T^{\mu\nu}u_\nu=\e u^\mu$. We work with the Minkowskian metric tensor $g^{\mu\nu}\equiv\mathrm{diag}(+,-,-,-)$.

The energy-momentum conservation, $\partial_\mu T^{\mu\nu} =0$, and particle four-current conservation, $\partial_\mu N^{\mu}=0$, yields the fundamental evolution equations for $\e$, $u^\mu$ and $n$, as
\begin{align}
\dot\e + (\e+\p+\Pi)\theta - \pi^{\mu\nu}\sigma_{\mu\nu} &= 0,  \label{evol1}\\
(\e+\p+\Pi)\dot u^\alpha - \nabla^\alpha (\p+\Pi) + \Delta^\alpha_\nu \partial_\mu \pi^{\mu\nu}  &= 0, \label{evol2}\\
\dot n + n\theta + \partial_\mu n^{\mu} &=0. \label{evol3}
\end{align}
Here we use the standard notation $\dot A=u^\mu\partial_\mu A$ for co-moving derivatives, $\nabla^\alpha=\Delta^{\mu\alpha} \partial_\mu$ for space-like derivatives, $\theta\equiv\partial_\mu u^\mu$ for the expansion scalar and, $\sigma^{\mu\nu}\equiv\frac{1}{2}(\nabla^\mu
u^\nu+\nabla^\nu u^\mu)-\frac{1}{3}\theta\Delta^{\mu\nu}$ for the velocity stress tensor.

In order to calculate the thermodynamic quantities corresponding to a system of single species of relativistic particles, we consider the equilibrium phase-space distribution functions for particles, $f_{\rm eq}$, and anti-particles, $\bar{f}_{\rm eq}$, given as:
\begin{equation}\label{feq}
\displaystyle{f_{\rm eq} \equiv \frac{1}{e^{\beta(u\cdot p)- \alpha} + a}} , \qquad  
\displaystyle{\bar{f}_{\rm eq} \equiv \frac{1}{e^{\beta(u\cdot p)+ \alpha} + a}} \, ,
\end{equation}
where $u\!\cdot\! p \equiv u_\mu p^\mu$, $\beta\equiv 1/T$ is the inverse temperature and $\alpha\equiv \mu/T$ is the ratio of the chemical potential to temperature. Also, $a$ represents the species of particles; $a=-1,0,1$  for Bose-Einstein, Boltzmann, and Fermi-Dirac gas, respectively.

The temperature, $T$, and chemical potential, $\mu$, of a non-equilibrium system are auxiliary quantities and are defined using the matching conditions $\e=\e_0$ and $n=n_0$, where $\e_0$ and $n_0$ are the energy density and the net number density in equilibrium:
\begin{align}
\e_0 &\equiv u_\mu u_\nu \!\int\! dP \, p^\mu p^\nu \left( f_{\rm eq}+\bar{f}_{\rm eq} \right) = I_{2,0}^{+} \ ,  \label{en_den} \\  
n_0 &\equiv u_\mu \!\int\! dP \, p^\mu \left( f_{\rm eq}-\bar{f}_{\rm eq} \right) = I_{1,0}^{-} \ .  \label{num_den}
\end{align}
Here, the thermodynamic integrals $I_{n,q}^{\pm}$ are defined as
\begin{equation}\label{Inq}
I_{n,q}^{\pm} \equiv \frac{1}{(2q+1)!!} \!\int \!\mathrm{dP} \left( u \!\cdot\! p \right)^{\!n-2q} \!\left( \Delta_{\alpha\beta}\, p^\alpha p^\beta \right)^{\!q} \! \left(f_{\rm eq} \pm \bar{f}_{\rm eq} \right) .
\end{equation}
It is easy to see that the equilibrium pressure and entropy density are 
\begin{align}
\p &\equiv -\frac{1}{3}\Delta_{\mu\nu} \!\int\! dP \, p^\mu p^\nu \left( f_{\rm eq}+\bar{f}_{\rm eq} \right) = -I_{2,1}^{+} \ , \label{prs} \\
s_0 &\equiv \frac{\e_0+\p-\mu\,n_0}{T} = \frac{1}{T}  \left( I_{2,0}^{+} - I_{2,1}^{+} -\mu\, I_{1,0}^{-} \right) . \label{ent_den}
\end{align}
Using the expressions for these thermodynamic quantities, the speed of sound squared can be obtained from $c_s^2 \equiv d\p/d\e\big|_{s/n}$. 

For a system close to local thermodynamic equilibrium, the non-equilibrium phase-space distribution function can be written as $f=f_{\rm eq}+\delta f$, where $|\delta f|\ll f$ is out-of-equilibrium correction to the distribution function. Using Eqs.~\eqref{cons_emt} and \eqref{cons_curr}, the bulk viscous pressure, $\Pi$, the shear stress tensor, $\pi^{\mu\nu}$ and the dissipative charge diffusion, $n^\mu$, are expressed in terms of $\delta f$ as
\begin{align}
\Pi &= -\frac{1}{3}\Delta_{\alpha\beta} \!\int\! dP \, p^\alpha p^\beta\, (\delta f +\delta \bar{f}), \label{BVP}\\
\pi^{\mu\nu} &= \Delta^{\mu\nu}_{\alpha\beta} \!\int\! dP \, p^\alpha p^\beta\, (\delta f+\delta \bar{f}), \label{SST}\\
n^\mu &= \Delta_\nu^\mu \!\int\! dP\, p^\nu (\delta f-\delta \bar{f}) ,\label{CD}
\end{align}
where $\Delta^{\mu\nu}_{\alpha\beta}\equiv \frac{1}{2}(\Delta^{\mu}_{\alpha}\Delta^{\nu}_{\beta} + \Delta^{\mu}_{\beta}\Delta^{\nu}_{\alpha}) - \frac{1}{3}\Delta^{\mu\nu}\Delta_{\alpha\beta}$ is a traceless symmetric projection operator orthogonal to $u^\mu$ as well as $\Delta^{\mu\nu}$. In order to derive the expressions for the above dissipative quantities in terms of hydrodynamic gradients and calculate the associated transport coefficients, we require the form of $\delta f$ and $\delta \bar{f}$. In the next section, we obtain the first-order correction to $f_{\rm eq}$ by extending the Anderson-Witting RTA approximation of the collision term to incorporate energy dependence of relaxation-time.

\section{Boltzmann equation} \label{sec:BE}

The evolution of single-particle phase-space distribution function $f(x,p)$ within the framework of kinetic theory is governed by the Boltzmann equation:
\begin{equation}\label{BE}
p^\mu \partial_\mu f = \mathcal{C}\left[f, \bar{f}\right], \qquad
p^\mu \partial_\mu \bar{f} = \bar{\mathcal{C}}\left[f, \bar{f}\right], 
\end{equation}
where $\mathcal{C}[f, \bar{f}]$ and $\bar{\mathcal{C}}[f, \bar{f}]$ are the so called collision kernel which is the most nontrivial part of Boltzmann equation. For a system close to equilibrium, Anderson and Witting proposed an approximation for the collision kernel,
\begin{align}
\mathcal{C}[f, \bar{f}] \!=\! - \frac{(u\!\cdot\! p)}{\taur(x)} (f - f_{\rm eq}) \, , \quad  
\bar{\mathcal{C}}[f, \bar{f}] \!=\! - \frac{(u\!\cdot\! p)}{\bar{\tau}_{\rm R}(x)} (\bar{f} - \bar{f}_{\rm eq}) \, , \label{AWRTA2}
\end{align}
where $\taur$ and $\bar{\tau}_{\rm R}$ are relaxation time scales for particles and anti-particles, respectively. The above model, which we refer to as RTA in the following, assumes $\taur$ and $\bar{\tau}_{\rm R}$ to be independent of particle energies. Conservation of energy-momentum tensor and net-charge current, defined in Eqs.~\eqref{cons_emt}~and~\eqref{cons_curr}, leads to
\begin{align}
\partial_\mu T^{\mu\nu} = \partial_\mu\!\int\! dP p^\mu p^\nu \left( f + \bar{f} \right) &= 0 \label{emt_cons} \\ 
\partial_\mu N^\mu = \partial_\mu\!\int\! dP\, p^\mu \left( f-\bar{f} \right) &= 0. \label{curr_cons}
\end{align}
This imposes the following conditions on the collision term in the RTA approximation defined in Eq.~\eqref{AWRTA2}:
\begin{align}
u_\mu \int  dP\, p^\mu p^\nu \left[ \frac{f}{\taur}+\frac{\bar{f}}{\bar{\tau}_{\rm R}} \right] &= u_\mu \int  dP\, p^\mu p^\nu \left[ \frac{f_{\rm eq}}{\taur}+\frac{\bar{f}_{\rm eq}}{\bar{\tau}_{\rm R}} \right], \label{cons_eq_1}\\
u_\mu \int  dP\, p^\mu  \left[ \frac{f}{\taur}-\frac{\bar{f}}{\bar{\tau}_{\rm R}} \right]  &= u_\mu \int  dP\, p^\mu  \left[ \frac{f_{\rm eq}}{\taur}-\frac{\bar{f}_{\rm eq}}{\bar{\tau}_{\rm R}} \right]. \label{cons_eq_2}
\end{align}
For particle energy independent relaxation times, the above two conditions are consistent with the Landau frame condition, $T^{\mu\nu}u_\nu=\e u^\mu$, and matching conditions, $\e=\e_0$ and $n=n_0$, provided that the relaxation time scales for particles and antiparticles are equal, i.e., $\taur=\bar{\tau}_{\rm R}$~\cite{Bhadury:2020ngq}. In the following we will assume this to be true and assign only one relaxation timescale, $\taur$, for the system under consideration. However, it can be easily seen that conservation  equations~\eqref{cons_eq_1}~and~\eqref{cons_eq_2} are not satisfied even in the Landau frame when the relaxation time $\taur$ becomes particle energy (or momentum) dependent.
Therefore the form of usual RTA is rather restrictive and in what follows, we extend the RTA such that the 
conservation equations are satisfied for energy-dependent relaxation time.

\subsection{Extended Relaxation time approximation} \label{sec:ERTA}

We consider the following extension of the RTA \cite{Teaney:2013gca}:
\begin{align}
p^\mu \partial_\mu f &= - \frac{(u\!\cdot\!p)}{\taur(x,p)} \left(f - f_{\rm eq}^{*}\right) ,\label{ERTA1}
\end{align}
with $f \to \bar{f}$ and $f_{\rm eq}^{*} \to \bar{f}_{\rm eq}^{*}$ for anti-paticles. Here, $f_{\rm eq}^{*}$ and $\bar{f}_{\rm eq}^{*}$ are the local equilibrium distribution functions upon which the system relaxes with a time scale $\taur(x, p)$,
\begin{equation}\label{feq*}
\displaystyle{f_{\rm eq}^{*} \equiv \frac{1}{e^{\beta^*(u^*\cdot p)- \alpha^*} + a}} , \quad  
\displaystyle{\bar{f}_{\rm eq}^{*} \equiv \frac{1}{e^{\beta^*(u^*\cdot p)+ \alpha^*} + a}} \, ,
\end{equation}
where $\beta^*\equiv 1/T^*$ and $\alpha^*\equiv \mu^*/T^*$. Note that the above equilibrium distribution functions reduces to the Maxwell-Boltzmann/Fermi-Dirac/Bose-Einstein form in the local rest frame of $u_\mu^*$ with temperature $T^*$ and chemical potential $\mu^*$. In this sense, we interpret the local rest frame of $u_\mu^*$ to be the `thermodynamic frame' \cite{Banerjee:2012iz, Jensen:2012jh, Kovtun:2019hdm} with $T^*$ and $\mu^*$ being the corresponding thermodynamic quantities. 

At this juncture, we would like to clarify the interpretation and differences between hydrodynamic frame and thermodynamic frame. Firstly, we emphasize that hydrodynamic ``frames''  (for instance the two well known Landau and Eckart frames or any other definition of $u^\mu$), should not be confused with Lorentz frames, i.e., they are not related to each other by Lorentz transformations. Instead, these definitions of $u^\mu$ specifies the local fluid rest frame. Moreover, for a non-equilibrium system, temperature and chemical potential are auxiliary fields which we define using the matching conditions. Therefore the local equilibrium distribution function, defined in Eq.~\eqref{feq}, is of the Maxwell-Boltzmann/Fermi-Dirac/Bose-Einstein form in the fluid rest frame, $u^\mu=(1,0,0,0)$, with auxiliary fields $T$ and $\mu$. On the kinetic theory side, where we consider the Boltzmann equation in ERTA [Eq.~\eqref{ERTA1}], we propose that the distribution function relaxes to a local equilibrium distribution, Eq.~\eqref{feq*}, which takes the form of Maxwell-Boltzmann/Fermi-Dirac/Bose-Einstein distributions in the rest frame of $u_\mu^*$. In this local rest frame definition, which we term as the thermodynamic frame, the distribution corresponds to temperature $T^*$ and chemical potential $\mu^*$. We insist that the local rest frame of $u^\mu$ (defined in the present work as the Landau frame) on the hydrodynamics side need not coincide with that of $u_\mu^*$ on the kinetic theory side. In the following, we will relate these two frames, and the corresponding thermodynamic quantities.

\subsection{Relation between two frames} \label{sec:2eq}

In order to obtain the out-of-equilibrium correction to the distribution function, we employ Chapman-Enskog like expansion to iteratively solve the ERTA Boltzmann equation~\eqref{ERTA1}, i.e., $f = f_{\rm eq} + \delta f_{(1)} + \delta f_{(2)} + \dots$. Here $\delta f_{(i)}$ represents $i^{\rm th}$ order gradient correction to the distribution function. Since we are interested in deriving hydrodynamic equations, the expansion is done about the hydrodynamic equilibrium. The first order gradient correction, $\delta f_{(1)}$ obtained using Eq.~\eqref{ERTA1}, 
\begin{equation}\label{delta_f1}
\delta f_{(1)} = \delta f_{*} - \frac{\taur(x,p)}{(u\cdot p)} \  p^\mu \partial_\mu f_{\rm eq},
\end{equation}
and similarly for $\delta \bar{f}_{(1)}$. Here, $\delta f_*\equiv f_{\rm eq}^{*} - f_{\rm eq}$ adds a further gradient correction to $\delta f_{(1)}$ arising from the difference between the definition of hydrodynamic and thermodynamic frame variables. It is clear that these `frames' should coincide in equilibrium (as the collision term vanishes in equilibrium), and therefore $\delta f_*$ vanishes in equilibrium.

Towards determining $\delta f_{*}$, we start by relating the auxiliary hydrodynamic variables $u^\mu, T$ and $\mu$ to the corresponding variables on the kinetic theory side, $u_\mu^*, T^*$ and $\mu^*$,
\begin{equation}\label{frame_rel}
u_\mu^* \equiv u_\mu + \delta u_\mu, \qquad T^* \equiv T + \delta T , \qquad \mu^* \equiv \mu + \delta \mu \ .
\end{equation}
Since $\delta u_\mu, \ \delta T$, and $\delta \mu$ are out of equilibrium corrections, they are at least first order in gradients%
     \footnote{Since both $u_\mu$ and $u^*_\mu$ are time like four-velocities, their normalization leads to, $u\cdot u = u^{*}\cdot  u^{*} = 1, \implies  u\cdot\delta u = \mathcal{O}\!\left(\delta^2\right).$ Hence the term $u\cdot\delta u$ is second-order 
     in gradients and we shall ignore it in the present derivation.}. 
To obtain $\delta f_{*}$, we Taylor expand $f_{\rm eq}^{*}$ about $u^\mu$, $T$ and $\mu$:
\begin{align}
\delta f_{*} &=  \frac{\partial f_{\rm eq}^{*}}{\partial u^{\mu}_{*}} \Bigg|_{(u^\mu, T, \mu)} \!\!\!\delta u^{\mu} \,+\, \frac{\partial f_{\rm eq}^{*}}{\partial T_{*}} \Bigg|_{(u^\mu, T, \mu)}\!\!\! \delta T 
+\, \frac{\partial f_{\rm eq}^{*}}{\partial \mu_{*}} \Bigg|_{(u^\mu, T, \mu)}\!\!\! \delta\mu \,+\, \mathcal{O}\!\left(\delta^2\right)
\nonumber \\
&= \left[- \frac{p_\mu \delta u^\mu}{T} + \frac{(u\!\cdot\!p - \mu)\delta T}{T^2}  + \frac{\delta \mu}{T} \right] f_{\rm eq} \tilde{f}_{\rm eq} + \mathcal{O}\!\left(\delta^2\right) ,
\label{delta_f*}
\end{align}
where $\tilde{f}_{\rm eq}\equiv 1-af_{\rm eq}$. For anti-particles, we replace the chemical potential $\mu\to -\mu$ and $f_{\rm eq} \to \bar{f}_{\rm eq}$. In deriving, we have ignored terms of $\mathcal{O}\!\left(\delta^2\right)$ as they are of higher order. 

\subsection{Out-of-equilibrium correction to the distribution function} \label{sec:CE_exp}

We state some useful relations and identities which will be used later to obtain the first-order correction, $\delta f_{(1)}$. Using the hydrodynamic equations~\eqref{evol1}-\eqref{evol3} and Eqs.~\eqref{en_den}, \eqref{num_den}, we obtain
\begin{align} 
\dot{\alpha} &= \chi_a\, \theta + \mathcal{O}(\delta^2) \,, \quad
\dot{\beta} = \chi_b\, \beta\,\theta + \mathcal{O}(\delta^2) \,, \label{alpha_beta_dot}\\ 
\nabla^\mu \beta &= \frac{n}{\e+\p} (\nabla^\mu \alpha) - \beta\, \dot{u}^\mu + \mathcal{O}(\delta^2) \,, \label{nabla_beta}
\end{align}
where we have kept all terms till first-order in gradients. The dimensionless quantities $\chi_a$ and $\chi_b$ are defined as
\begin{equation}
\chi_a \equiv \frac{(\e\!+\!\p) J_{2,0}^{-} -n J_{3,0}^{+} }{J_{3,0}^{+} J_{1,0}^{+} - J_{2,0}^{-} J_{2,0}^{-}} , \quad
\chi_b \equiv \frac{(\e\!+\!\p) J_{1,0}^{+} -n J_{2,0}^{-} }{\beta(J_{3,0}^{+} J_{1,0}^{+} - J_{2,0}^{-} J_{2,0}^{-})}.
\end{equation}
The $J_{n,q}^{\pm}$ integrals appearing in the above expressions are defined similar to Eq.~\eqref{Inq}, but with $(f_{\rm eq} \Tilde{f}_{\rm eq} \pm \bar{f}_{\rm eq} \Tilde{\bar{f}}_{\rm eq})$ in the integrand instead of $(f_{\rm eq} \pm \bar{f}_{\rm eq} )$.

Using Eqs.~\eqref{delta_f*}, \eqref{alpha_beta_dot} and \eqref{nabla_beta}, the first-order gradient correction to the distribution function, Eq.~\eqref{delta_f1}, simplifies to
\begin{align}\label{delta_f1_exp}
\delta f_{(1)} =& \left[ -\beta\, p\!\cdot\!\delta u + \beta^2 \left( u\!\cdot\! p - \mu \right) \delta T + \beta\, \delta \mu \right] f_{\rm eq} \, \tilde{f}_{\rm eq}  \nonumber \\
& +\taur(x,p) \left[ \left( \beta (u\!\cdot\! p) \left(\chi_b-1/3\right) 
+ \frac{\beta\, m^2}{3(u\!\cdot\! p)} - \chi_a \right) \theta \right. \nonumber \\
& + \left. \frac{\beta}{u\!\cdot\! p}\, p^\mu p^\nu \sigma_{\mu\nu}  
+ \left( \frac{n}{\e\!+\!\p} - \frac{1}{u\!\cdot\! p} \right) p^\mu \nabla_\mu \alpha \right] f_{\rm eq} \, \tilde{f}_{\rm eq} .
\end{align}
To evaluate $\delta\bar{f}_{(1)}$ for anti-particles, one needs to replace $\mu\!\to\!-\mu$, $\alpha\!\to\!-\alpha,~\chi_a\!\to\!-\chi_a$ and $f_{\rm eq} \,\tilde{f}_{\rm eq}\to \bar{f}_{\rm eq}\,\tilde{\bar{f}}_{\rm eq}$ in above expression. 

\subsection{Hydrodynamic frame and matching conditions} \label{sec:frame_match}

We note that $\delta f_{(1)}$ depends on $\delta T$, $\delta u^\mu$ and $\delta \mu$. Imposing Landau frame conditions: $u_\nu T^{\mu \nu} = \e u^\mu$ and the matching conditions: $\e = \e_0$  and $n= n_0$ at first-order in gradients, i.e., with $f=f_1\equiv f_{\rm eq}+ \delta f_{(1)}$, we obtain
\begin{equation}\label{deluTmu}
\delta u^\mu = \mathcal{C}_1 \frac{(\nabla^\mu \alpha)}{T}  \,, \qquad
\delta T = \mathcal{C}_2\, \theta \,, \qquad 
\delta \mu = \mathcal{C}_3\, \theta \,,
\end{equation}
where we have introduced the dimensionless variables
\begin{align}
\mathcal{C}_1 &\equiv \frac{T}{\e+\p}\left[ K_{2,1}^{-} - \left(\frac{n}{\e+\p} \right) K_{3,1}^{+} \right], \label{C1}\\
\mathcal{C}_2 &\equiv T^2\left(\frac{ J_{2,0}^- \ y_{2,0}^{-} - J_{1,0}^+ \ y_{3,0}^{+} }{J_{3,0}^+ J_{1,0}^+ - J_{2,0}^- J_{2,0}^- }\right), \label{C2}\\
\mathcal{C}_3 &\equiv T \left[ \frac{ \left( J_{2,0}^- -\mu\, J_{1,0}^+\right) y_{3,0}^{+} - \left( J_{3,0}^+ -\mu\, J_{2,0}^-\right) y_{2,0}^{-}  }{J_{3,0}^+ J_{1,0}^+ - J_{2,0}^- J_{2,0}^- }\right]. \label{C3}
\end{align}
In the above equations, $K_{n,q}^{\pm}$ are defined similar to Eq.~\eqref{Inq} but with $\left[(f_{\rm eq}\Tilde{f}_{\rm eq} \pm \bar{f}_{\rm eq} \Tilde{\bar{f}}_{\rm eq})\taur(x,p)\right]$ in the integrand instead of $(f_{\rm eq} \pm \bar{f}_{\rm eq} )$. The coefficients $y_{n,q}^{\pm}$ are defined as
\begin{align}
    y_{n,q}^{\pm} \equiv \beta \left( \chi_b-\frac{1}{3} \right) K_{n,q}^{\pm} + \frac{m^2}{3T} K_{(n-2),q}^{\pm} - \chi_a K_{(n-1),q}^{\mp} \,.
\end{align}
When the relaxation time is particle energy independent, the integrals $K_{n,q}^{\pm} \to \taur(x) J_{n,q}^{\pm}$ and $y_{2,0}^{-} \,, y_{3,0}^{+}\to 0$ leading to vanishing of the coefficients $\mathcal{C}_1$, $\mathcal{C}_2$ and $\mathcal{C}_3$. Thus $\delta u^\mu ,\ \delta T$ and $\delta\mu$ vanishes and ERTA reduces to the usual Anderson-Witting RTA.

As already mentioned, the quantities $\delta u^\mu$, $\delta T$ and $\delta \mu$ relate the hydrodynamic and thermodynamic frames, and to express these in terms of hydrodynamic fields requires one to impose the hydrodynamic frame and matching condition at each order in gradients. Further, the derivation of $\delta f_{(1)}$ requires the lower order evolution of hydrodynamic fields, Eqs.~\eqref{alpha_beta_dot}~and~\eqref{nabla_beta}, obtained from conservation equations $\partial_\mu T^{\mu\nu}=0$ and $\partial_\mu N^\mu=0$ \textit{in the same hydrodynamic frame}. Therefore obtaining $\delta f_{(1)}$ by iteratively solving the ERTA Boltzmann equation, Eq.~\eqref{ERTA1}, and fixing $\delta T$, $\delta u^\mu$ and $\delta \mu$ using the same frame and matching conditions ensures macroscopic conservation by construction.

\section{First order transport coefficients} \label{sec:hydro_NS}

Equation~\eqref{delta_f1_exp} together with Eq.~\eqref{deluTmu} and Eqs.~\eqref{C1}-\eqref{C3} completely specifies the first-order correction to the equilibrium distribution function, $\delta f_{(1)}$. With this, the relativistic Navier-Stokes expression for dissipative quantities using the definitions given in Eqs.~\eqref{BVP}-\eqref{CD} is obtained to be,
\begin{equation}\label{rel_NS}
\pi^{\mu\nu} = 2\, \eta\, \sigma^{\mu\nu}, \qquad
\Pi = - \zeta\, \theta, \qquad
n^\mu = \kappa_n \nabla^\mu \alpha.
\end{equation}
where the transport coefficients are given by,
\begin{align}
\eta &= \frac{K_{3,2}^{+}}{T} \,, \label{shear_visc} \\ 
\zeta &= -\mathcal{C}_2 \left(\frac{ \e+\p - \mu\, n }{T}\right) - \mathcal{C}_3\, n + y_{3,1}^{+} \,,
\label{bulk_visc} \\
\kappa_n &= \mathcal{C}_1\, \frac{n}{T} + \left( \frac{n}{\e+\p} \right) K_{2,1}^{-} - K_{1,1}^{+} \,. \label{conductivity}
\end{align}

The form of the relaxation time $\tau_R(x,p)$ has not yet been specified. In the next Section, we study the behaviour of transport coefficients using a power-law parametrization for the energy dependence of the relaxation time \cite{Dusling:2009df, Chakraborty:2010fr, Dusling:2011fd, Teaney:2013gca, Kurkela:2017xis},
\begin{align}\label{param}
\taur (x,p)= \taueq(x)\, \left( \frac{u\cdot p}{T} \right)^{\ell} \,.
\end{align}
Here $\taueq(x)$ represents the energy independent part of relaxation time and $\ell$ is a constant. 
In this case, the integrals $K_{n,q}^{\pm}$ can be expressed as: $K_{n,q}^{\pm} = \left(\taueq/T^\ell\right)\, J_{(n+\ell),q}^{\pm}$ . 

\section{Results and discussions}
\label{sec:res_disc}

In this section, we study the behaviour of first-order transport coefficients obtained in the Section~\ref{sec:hydro_NS} for three different cases:
\begin{enumerate}
\item Chargeless and massless particles ($m=\mu=0$): bulk viscous pressure and charge flux vanishes and dissipation in the system occurs purely from shear stress tensor. 
\item Chargeless massive particles ($\mu=0$ and $m\neq0$): charge flux vanishes and dissipation in the system is due to shear stress tensor and bulk viscous pressure.
\item Charged massless particles ($m=0$ and $\mu\neq0$): bulk viscous pressure vanishes and dissipation in the system is due to non-zero shear stress tensor and charge flux. 
\end{enumerate}

\subsection{System of chargeless and massless particles}
\label{sec:etabys}

A system of massless ($m=0$) and chargeless ($\mu=0$) particles has dissipation purely due to shear stress tensor. For such a system, the only relevant transport coefficient is the coefficient of shear viscosity. Since $\mu=0$, the integrals $I_{n,q}^-=J_{n,q}^-=0$, and the coefficients $\mathcal{C}_1,\,\mathcal{C}_2,\,\mathcal{C}_3$ vanishes. 
The coefficient of shear stress given by  Eq.~\eqref{shear_visc} reduces to,
\begin{equation}\label{shear_ml_cl}
\eta =  \frac{\taueq\,g\,T^4\,\Gamma(5+\ell)}{15\,\pi^2}\!
\left[\frac{\mathrm{Li}_{4+\ell}(-a)}{-a}\right],
\end{equation}
where $g$ is the degeneracy factor and $a=0$, $-1$, and $1$ corresponds to Maxwell-Boltzmann (MB), Bose-Einstein (BE) and Fermi-Dirac (FD) statistics, respectively. $\Gamma(x)$ is the Euler Gamma function and $\mathrm{Li}_n(x)\equiv\sum_{k=1}^{\infty}(x^k/k^n)$ is the polylogarithm (also referred to as Jonqui\'ere's function). Note that the the quantity inside the square-brackets on the right hand side of the above equation is positive definite for all $\ell$ and is equal to $1$, $\zeta(4+\ell)$ and $(1-2^{-3-\ell})\zeta(4+\ell)$ for $a=0$, $-1$, and $1$, respectively with $\zeta(s)\equiv\sum_{k=1}^{\infty}(1/k^s)$ being the Riemann zeta function. 

\begin{figure}[t!]
\centering
 \includegraphics[width=.95\linewidth]{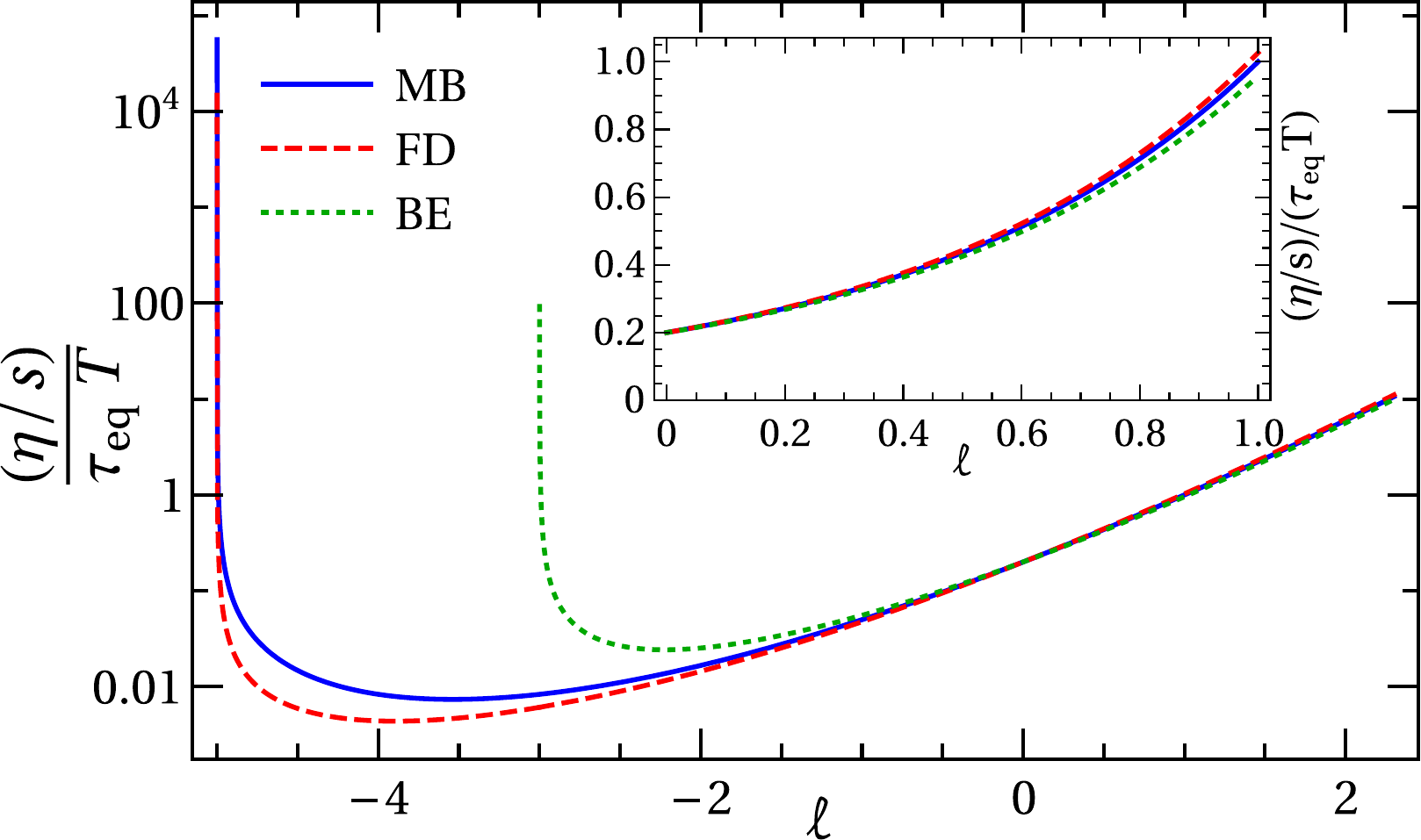}
 \vspace{-2mm}
 \caption{Dependence of $\eta/(s\taueq T)$ on $l$ for Maxwell-Boltzmann (MB), Fermi-Dirac (FD) and Bose-Einstein (BE) distributions. Inset shows the same within the range $0\leq\ell\leq 1$, relevant for QCD system.
	\label{fig:etabys_vs_l}}
	 \vspace{-2mm}
\end{figure}

To study the relative importance of shear viscosity, it is instructive to consider its dimensionless ratio with entropy density, $\eta/s$. 
In the conformal chargeless case, the out-of-equilibrium entropy density is the same as that in equilibrium, $s=s_0$ ($s_0$ is given in Eq.~\eqref{ent_den}), for first-order dissipative hydrodynamics \cite{Jaiswal:2013fc}.
The only energy scale is temperature in this case, and therefore $\taueq\sim1/T$. The quantity $\eta/(s\taueq T)$ is then just a function of $\ell$ which can be expressed using Eqs.~\eqref{ent_den}~and~\eqref{shear_ml_cl} as
\begin{equation}\label{shear_visc_ent_den}
\frac{\eta}{s\taueq T} = \frac{\Gamma(5+\ell)}{120}
\left[\frac{\mathrm{Li}_{4+\ell}(-a)}{\mathrm{Li}_{4}(-a)} \right].
\end{equation}
We observe that $\eta/(s\taueq T)$ for $\ell=0$ is independent of statistics. This follows since $s_0= -J_{3,1}^{+}/T^2$, and $\eta=\tau_{\rm eq} J_{3,2}^{+}/T$ when relaxation time is particle energy independent. Also, for massless particles, the integral $J_{3,2}^{+}=- J_{3,1}^{+}/5$, which leads to cancellation of the integrals in $\eta/(s\taueq T)$ and it takes the value $1/5$ independent of statistics. We see this feature in Fig.~\ref{fig:etabys_vs_l} where the three curves for MB, FD and BE cross each other at $\ell=0$.

We also observe a minimum in $\eta/(s\taueq T)$ for negative values of $\ell$ in Fig.~\ref{fig:etabys_vs_l} which occurs at different values of $\ell$ for different equilibrium statistics; $\ell\simeq-3.5,\,-3.9$ and $-2.2$ for MB, FD and BE, respectively. This may be attributed to the fact that the positive (negative) sign of $\ell$ results in the decrease (increase) of interaction strength with particle energy. A more detailed analysis of this feature is left for future work. In the inset, we focus on the range $0\leq \ell\leq 1$ which is relevant for QCD medium \cite{Dusling:2009df}. We observe that the specific viscosity increases with increasing $\ell$, having negligible dependence on the form of equilibrium statistics.

\subsection{System of chargeless massive particles}
\label{sec:zetabyeta}

For a system consisting of chargeless ($\mu=0$) but massive ($m\neq0$) particles, the dissipation is due to shear stress tensor as well as bulk viscous pressure. In this case the integrals  $I_{n,q}^-=J_{n,q}^-=0$ and $\mathcal{C}_1, \mathcal{C}_3$ vanishes irrespective of $\ell$. The transport coefficients $\eta$ and $\zeta$ in Eqs.~\eqref{shear_visc}~and~\eqref{bulk_visc}, reduces to
\begin{align}
\eta &= \frac{\tau_{\rm eq}}{T^{\ell+1}} J_{3+\ell,2}^{+} \,,  \label{eta_mu0}  \\
\zeta &= -\mathcal{C}_2 \left(\frac{\e+\p}{T}\right) +  \frac{\tau_{\rm eq}}{T^{\ell+1}}\left(\frac{5}{3} J_{3+\ell,2}^{+} + c_s^2 J_{3+\ell,1}^{+} \right) . \label{zeta_mu0}
\end{align}
The coefficient $\mathcal{C}_2$ and $c_s^2$ are given by
\begin{align}
\mathcal{C}_2 &= -\frac{\tau_{\rm eq}}{T^{\ell-1}} \left(\frac{c_s^2 J_{3+\ell,0}^{+} + J_{3+\ell,1}^{+}}{J_{3,0}^{+}} \right), \qquad
c_s^2 = \chi_b = \frac{\e+\p}{\beta\, J_{3,0}^{+}}.
\end{align}
Here $z\equiv m/T$ is defined as the ratio of particle mass and temperature. Note that $\mathcal{C}_2$ vanishes when the relaxation time is energy independent, i.e., for $\ell=0$. The above expression for the squared speed of sound, $c_s^2$, agrees with Ref.~\cite{Jaiswal:2014isa} and \cite{Florkowski:2015lra} for classical and quantum statistics, respectively.

\begin{figure}[t!]
\centering
\includegraphics[width=.95\linewidth]{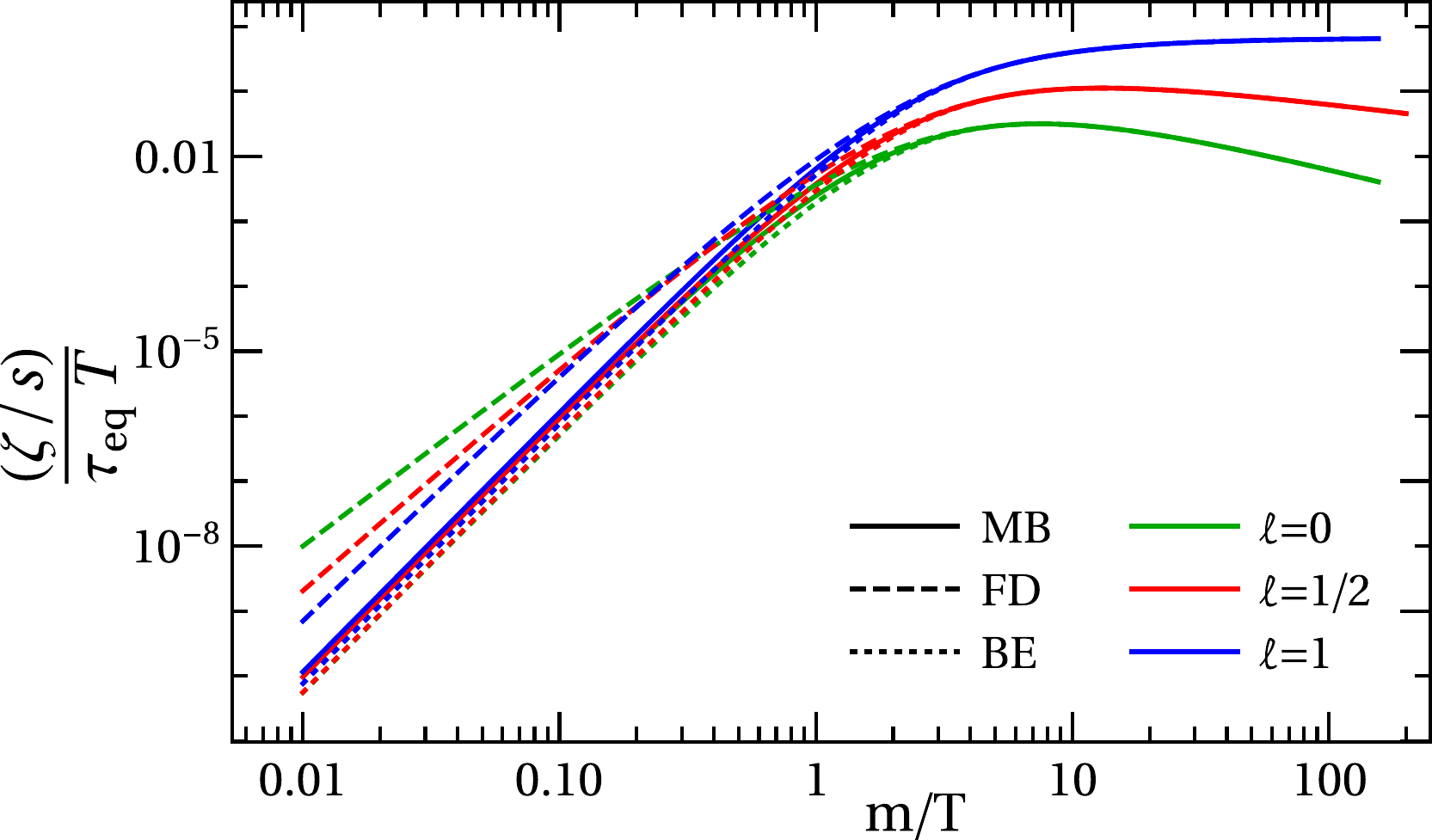}
\vspace{-2mm}
\caption{Variation of $\zeta/(s\taueq T)$ with $m/T$ for Maxwell-Boltzmann (MB), Fermi-Dirac (FD) and Bose-Einstein (BE) distributions for three different values of $\ell$.
\label{fig:zetabys_z} }
\vspace*{-2mm}
\end{figure}

In Fig.~\ref{fig:zetabys_z}, we show the variation of $\zeta/(s\taueq T)$ with $m/T$ for Maxwell-Boltzmann, Bose-Einstein and Fermi-Dirac distributions for three different values of $\ell$:  $0$, $1/2$ and $1$. We see a non-monotonous dependence on $m/T$ for all distributions and all values of $\ell$ considered here. This is in qualitative agreement with results obtained in Refs.~\cite{Mitra:2020gdk, Rocha:2021zcw}. We see that the present formulation does not lead to negative values for bulk viscosity, and therefore it does not violate  second law of thermodynamics. We now study the limiting behavior of the ratio of bulk viscosity to shear viscosity, $\zeta/\eta$.

\begin{figure*}[t!]
\includegraphics[width=\linewidth]{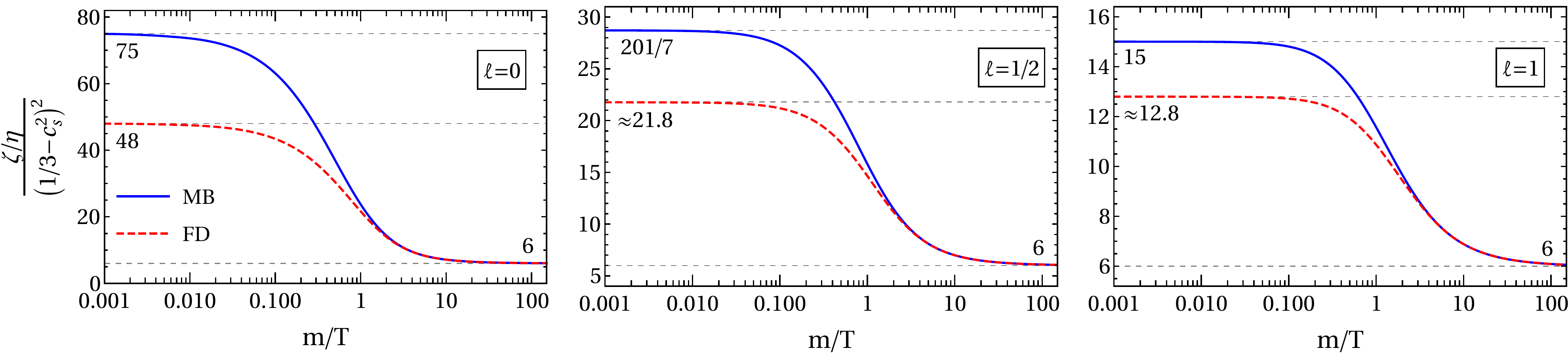}
\vspace*{-6mm}  
\caption{Variation of $\frac{\zeta/\eta}{\left(1/3-c_s^2\right)^2}$ with $m/T$ for Maxwell-Boltzmann (MB) and Fermi-Dirac (FD) statistics, for $\ell=0,~1/2$ and $1$.
\label{fig_Gammavsz} }
\vspace*{-4mm}
\end{figure*}

\textbf{Small $\mathbf{m/T}$ behavior:} 
It is instructive to study the scaling behavior of the ratio of viscous coefficients, $\zeta/\eta$, with the conformality measure, $1/3-c_s^2$. Small departure from conformality, i.e., small-$z\equiv m/T$ expansion for $1/3-c_s^2$ is given by
%
\begin{align}\label{cs2_small_z_exp}
\frac{1}{3}-c_s^2 = 
\begin{cases}
\frac{z^2}{36} +\mathcal{O}(z^3) \qquad~~\, \text{MB}\,, \\
\frac{5 z^2}{21 \pi ^2}+\mathcal{O}(z^3) \qquad \text{FD}\,, \\
\frac{5z^2}{12\pi^2}+\mathcal{O}(z^3) \qquad \text{BE}\,. 
\end{cases}
\end{align}
For MB and FD statistics, the quantity $\zeta/\eta$ in small-$z$ limit has the leading behavior as
\begin{align}\label{zetabyeta_MBFD}
\frac{\zeta}{\eta} = \Gamma \left(\frac{1}{3}-c_s^2 \right)^{2} \,,
\end{align}
where $\Gamma\equiv\lim\limits_{z\to0}\frac{\zeta/\eta}{\left(\frac{1}{3}-c_s^2 \right)^{2}}$. 
For classical MB statistics, we obtain
\begin{equation}\label{Gamma_MB}
\Gamma_{\rm MB} = \frac{15 \left(\ell^3+6 \ell^2-13 \ell+30\right)}{(\ell+1) (\ell+2) (\ell+3)},
\end{equation}
for positive values of $\ell$ which results in $\Gamma_{\rm MB}=75,~201/7$ and $15$ for $\ell=0,~1/2$ and $1$, respectively. For Fermi-Dirac statistics, we get analytical expressions only for integer $\ell$. We obtain $\Gamma_{\rm FD}=48$ for $\ell=0$ and $15+\frac{14 \left[7 \pi ^4 \log(2) -45 \pi ^2 \zeta (3)\right]}{375 \zeta (5)} \approx 12.8$ for $\ell=1$. For non-integer $\ell$, the value of this coefficient is obtained numerically and we get $\Gamma_{\rm FD}\simeq 21.8$ for $\ell=1/2$. These limiting values for MB and FD can be seen in Fig.~\ref{fig_Gammavsz} at $m/T\to 0$ slice.

\begin{figure}[b!]
\vspace*{-4mm} 
\begin{center}
\includegraphics[width=0.75\linewidth]{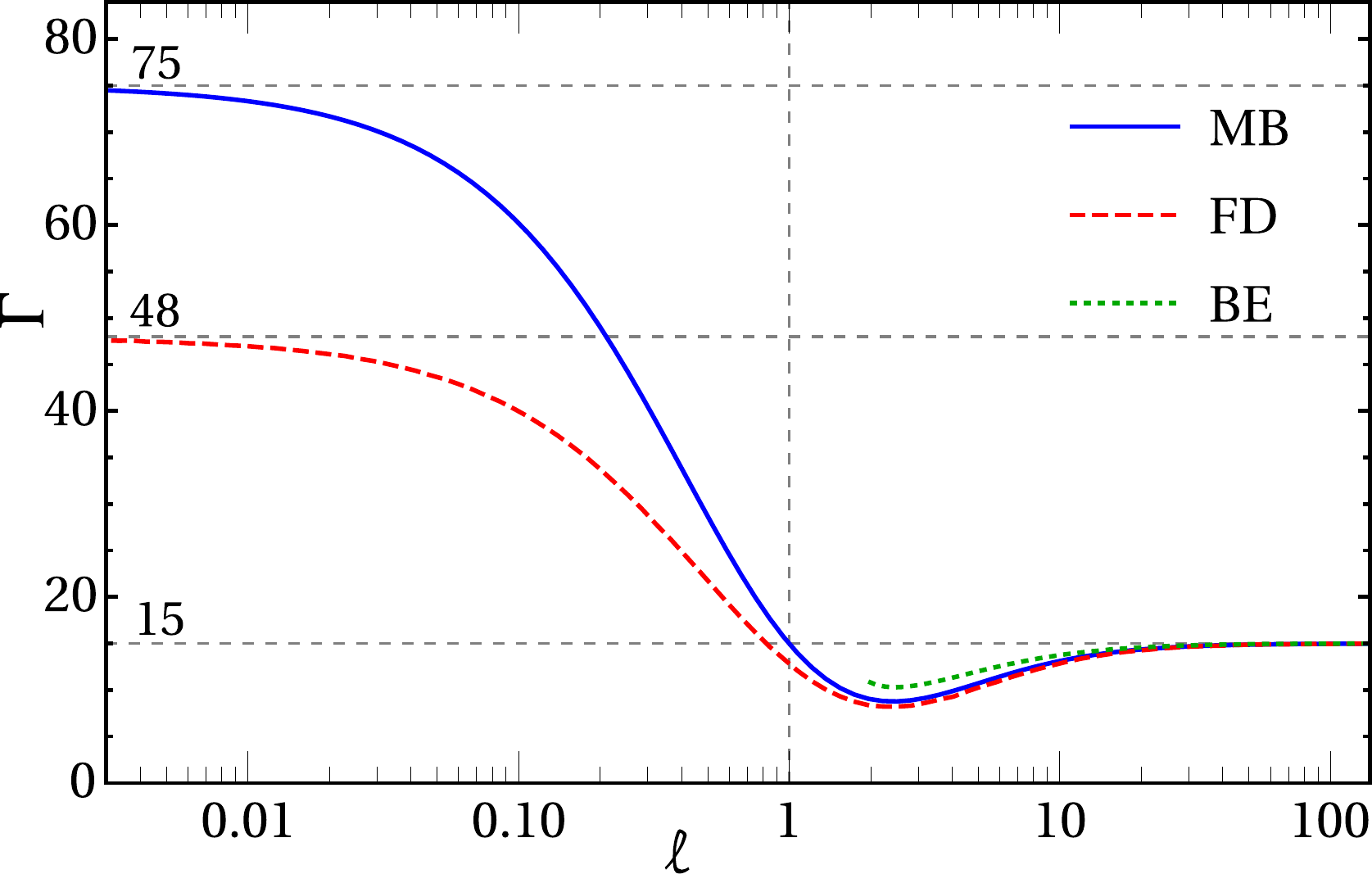}
\end{center}
\vspace*{-6mm}  
\caption{Behavior of $\Gamma$, defined below Eq.~\eqref{zetabyeta_MBFD}, with $\ell$ for Maxwell-Boltzmann (MB), Fermi-Dirac (FD) and Bose-Einstein (BE) equilibrium statistics.}
\label{fig_gamma_FDBE}
\end{figure} 

We observe a non-monotonic behavior of $\Gamma$ as a function of $\ell$ with a minimum at $\ell\approx 2.5$ in Fig.~\ref{fig_gamma_FDBE}. We also see that $\Gamma\to15$ for large $\ell$ irrespective of equilibrium statistics and converges to the MB case. This is the consequence of the strong dependence of the relaxation rate on the particle energy ($\taur \propto (u\cdot p)^\ell$) which governs the relaxation of low momentum particles towards equilibrium. Mathematically, this feature can be  understood from the following expression of the integrals for quantum statistics which can be written as a sum over Boltzmann factor~\cite{Florkowski:2015lra}:
\begin{align}
I_{n+\ell,q} =&\ \frac{gT^{n+\ell+2}z^{n+\ell+2}}{2\pi^2(2q+1)!!}(-1)^q\!\sum_{r=1}^{\infty}(-a)^{r-1}
\!\!\!\int_0^\infty\!\!\!\! d\theta\, \label{Inqr} \\
&\times(\cosh\theta)^{n+\ell-2q}(\sinh\theta)^{2q+2}\,\exp(-rz\cosh\theta), \nonumber\\
J_{n+\ell,q} =&\ \frac{gT^{n+\ell+2}z^{n+\ell+2}}{2\pi^2(2q+1)!!}(-1)^q\!\sum_{r=1}^{\infty}r(-a)^{r-1}
\!\!\!\int_0^\infty\!\!\!\! d\theta\, \label{Jnqr} \\
&\times(\cosh\theta)^{n+\ell-2q}(\sinh\theta)^{2q+2}\,\exp(-rz\cosh\theta). \nonumber
\end{align}
In above equations, for $\ell\gg1$ and $n+\ell\gg q$, the leading term, i.e., $r=1$ dominates and the other terms in the summation can be neglected. The integrals  $I_{n+\ell,q}$ and $J_{n+\ell,q}$ then reduces to that with classical MB statistics.  We find that a system with BE equilibrium statistics also follows the scaling relation of Eq.~\eqref{zetabyeta_MBFD} for large $\ell$ ($\gtrsim2$). At this point, it is worth mentioning that the scaling behavior in Eq.~\eqref{zetabyeta_MBFD} with $\Gamma=15$ was obtained by Weinberg for a system of radiation interacting with matter \cite{Weinberg:1971mx, Weinberg:1972kfs}.

The case of BE statistics for lower values of $\ell \, (< 2)$  is more subtle as the soft momenta governs the behavior of $\zeta/\eta$ in the small mass limit.
We follow the procedure outlined in Ref.~\cite{Czajka:2017wdo} to evaluate the thermodynamic integrals. For $\ell=0$, the leading term in small-$z$ expansion leads to the scaling relation
\begin{align}\label{zetabyeta_BE} 
\frac{\zeta}{\eta} = \frac{3\sqrt{15}}{2} \left(\frac{1}{3}-c_s^2 \right)^{3/2} \,.
\end{align}
As mentioned earlier, for $\ell\gtrsim2$, we find that a system with BE equilibrium statistics follows the scaling relation of Eq.~\eqref{zetabyeta_MBFD}, and the value of $\Gamma_{\rm BE}$ for quadratic scaling is plotted in Fig.~\ref{fig_gamma_FDBE}. The scaling behavior of $\zeta/\eta$ in the small-$z$ regime for $0<\ell<2$ is more involved and is left for future work.

\textbf{Large $\mathbf{m/T}$ behavior:} For constitutive particles having large mass compared to the temperature (non-relativistic regime), expansion of $c_s^2$ in powers of $1/z$ leads to
\begin{align}\label{cs2_large_m}
\frac{1}{3}-c_s^2 = \frac{1}{3}-\frac{1}{z} +\frac{3}{z^2}+\mathcal{O}\left(\frac{1}{z^3}\right)
\end{align}
which is independent of statistics. This is not surprising as for large mass, $a=\pm1$ in Eqs.~\eqref{feq}~and~\eqref{feq*} (which is relevant for quantum statistics) can be ignored and hence the thermodynamics is dominated by classical physics. Therefore the large mass limit is both non-relativistic and classical limit. Moreover, the ratio $\p/\e \to 0$ as $m/T\to \infty$ and hence $c_s^2$ vanishes. 

The quantity $\zeta/\eta$ in this limit has the expansion
\begin{align}\label{large_m_exp}
\frac{\zeta}{\eta} = \frac{2}{3}-\frac{4}{z} +\frac{26+(\ell-6) \ell}{z^2} + \mathcal{O}\left(\frac{1}{z^3}\right) \,.
\end{align}
Considering terms till $\mathcal{O}\left(\frac{1}{z}\right) $ in Eqs.~\eqref{cs2_large_m}~and~\eqref{large_m_exp}, we have
\begin{align}
\frac{\zeta}{\eta} = 6 \left( \frac{1}{3}-c_s^2 \right)^2,
\end{align}
which is irrespective of statistics as well as independent of `$\ell$', as can be seen in Fig.~\ref{fig_Gammavsz} for large values of $m/T$.

\subsection{System of charged massless particles} 
\label{sec:kappabyeta}

For a system of massless ($m=0$) particles with conserved charges ($\mu\neq0$), the dissipation is due to shear stress tensor and dissipative charge current. The expressions for shear viscosity, $\eta$, and charge conductivity, $\kappa_n$, are given in Eqs.~\eqref{shear_visc}~and~\eqref{conductivity}, respectively. 
In the following, we consider MB and FD statistics for analysis of charge and heat conduction. Calculation of charge/heat conductivity at finite chemical potential is not possible for Bose-Einstein statistics due to the phenomena of Bose-Einstein condensation.

For MB statistics, the ratio $\kappa_n/\eta$ in both small and large $\alpha (\equiv \mu/T)$ limit scales as 
\begin{equation}\label{kappn_eta}
\frac{\kappa_n}{\eta} = \Lambda_{\rm MB}\,\frac{1}{T},
\end{equation}
where the coefficient $\Lambda_{\rm MB}$ is a function of $\ell$, and for positive $\ell$ is given by
\begin{align}\label{Lambda_MB}
\Lambda_{\rm MB}=
\begin{cases}
\frac{5}{(4+\ell)(3+\ell)} \qquad &{\rm for}~\alpha\to0\,,\\
\frac{5(\ell^2-\ell+4)}{16\,(4+\ell)(3+\ell)} \qquad &{\rm for}~\alpha\to\infty\,.
\end{cases}
\end{align}
%

\begin{figure}[t!]
\begin{center}
\includegraphics[width=0.48\textwidth]{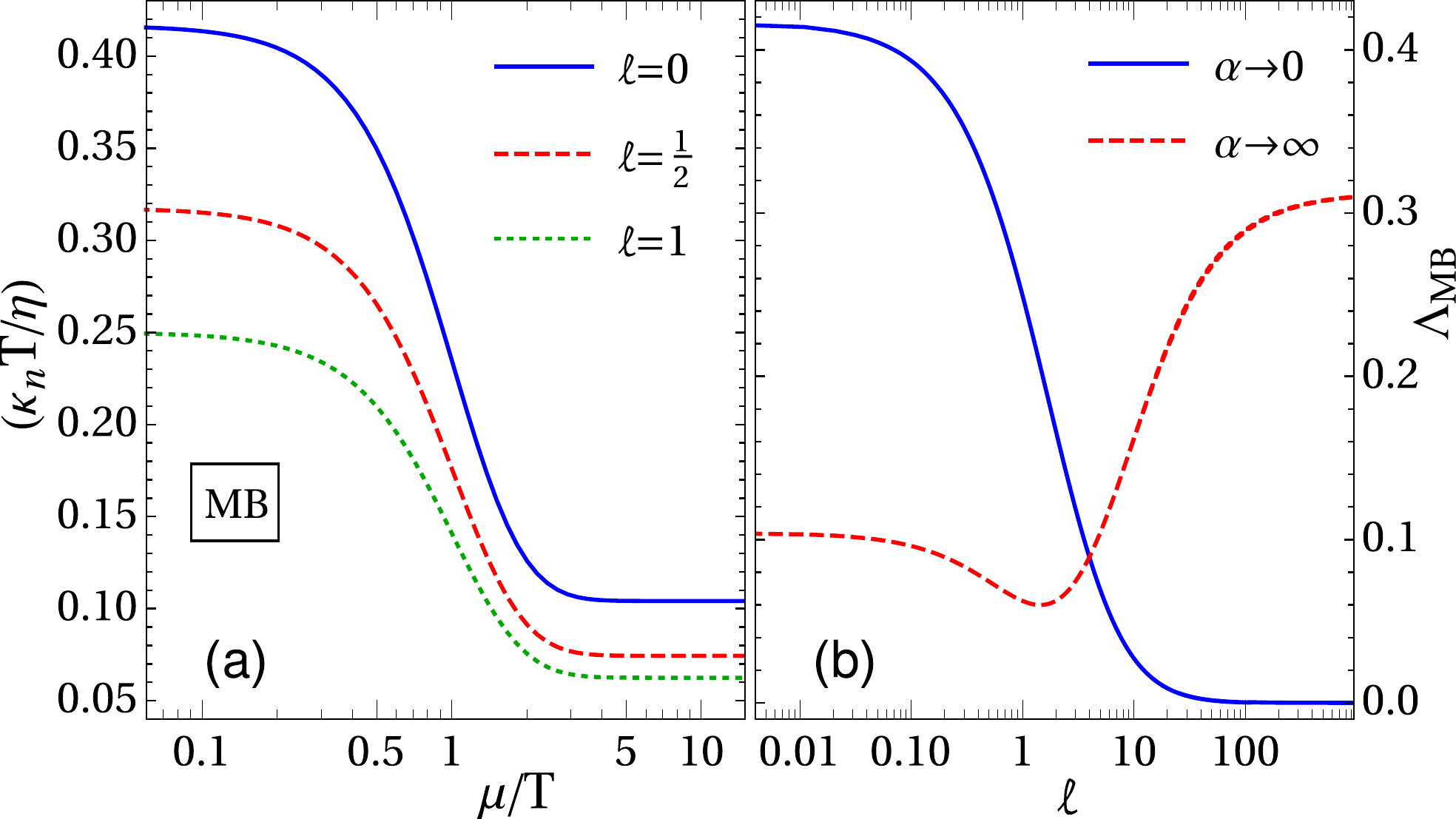}
\end{center}
\vspace*{-6mm}  
\caption{Maxwell-Boltzmann statistics: (a) Numerical evolution of the ratio of charge conductivity to shear viscosity multiplied with temperature as a function of $\mu/T$ for $\ell=0,~1/2$ and $1$. (b) The scaling coefficient $\Lambda_{\rm MB}$ defined in Eq.~\eqref{Lambda_MB} (for $\mu/T\to0$ and $\mu/T\to\infty$) as a function of $\ell$.}
\label{Lambda_MB_alpha}
\vspace*{-.4cm}
\end{figure} 

Fig.~\ref{Lambda_MB_alpha}a shows the evolution of the ratio 
$\kappa_n T/\eta$ as a function of $\mu/T$ for the case of MB statistics for $\ell=0,~1/2$ and $1$. As already mentioned, it can be seen that this ratio tends to constant values in the limit of both small and large $\mu/T$, indicating the scaling behavior in Eq.~\eqref{kappn_eta}. Further, we observe that this ratio is lower for increasing values of $\ell$. The solid blue and dashed red curves in Fig.~\ref{Lambda_MB_alpha}b shows the evolution of $\Lambda_{\rm MB}$ with $\ell$ in the limits $\alpha\to0$ and $\alpha\to\infty$, respectively, as given by Eq.~\eqref{Lambda_MB}. For both these solutions, we observe that $\Lambda_{\rm MB}$ saturates to constant values at both very small and very large values of $\ell$, indicating no $\ell$-dependence of $\Lambda_{\rm MB}$ in these limits. It is interesting to note the monotonous decrease of $\Lambda_{\rm MB}$ with $\ell$ for the solid blue curve, starting from $\Lambda_{\rm MB}=5/12$ for $\ell=0$ and approaching zero as $\ell\to \infty$.

For FD statistics, it is instructive to study the ratio $\kappa_q/\eta$, where $\kappa_q$ is the coefficient of thermal conductivity which can be written as $\kappa_q= \kappa_n \left(\frac{\e+\p}{nT}\right)^2$. We find that the ratio in both small and large $\mu/T$ limit scales as \cite{Son:2006em, Jaiswal:2015mxa}
\begin{equation}\label{kappq_eta}
\frac{\kappa_q}{\eta} = \Lambda_{\rm FD}\,\frac{\pi^2 T}{\mu^2},
\end{equation}
where the coefficient $\Lambda_{\rm FD}$ is again a function of $\ell$, and for positive values of $\ell$ in this two limits is given by
\begin{equation}\label{Lambda_FD}
\Lambda_{\rm FD}=
\begin{cases}
\frac{196\,\pi^2\left(2^{1+\ell}-1\right)\,\zeta(2+\ell)}{45\,\left(2^{3+\ell}-1\right)(4+\ell)(3+\ell)\zeta(4+\ell)} \quad &{\rm for}~\alpha\to0\,,\\
\frac{5}{3} \quad &{\rm for}~\alpha\to\infty\,.
\end{cases}
\end{equation}
where $\zeta(s)$ is the Riemann zeta function.

\begin{figure}[t!]
\begin{center}
\includegraphics[width=0.465\textwidth]{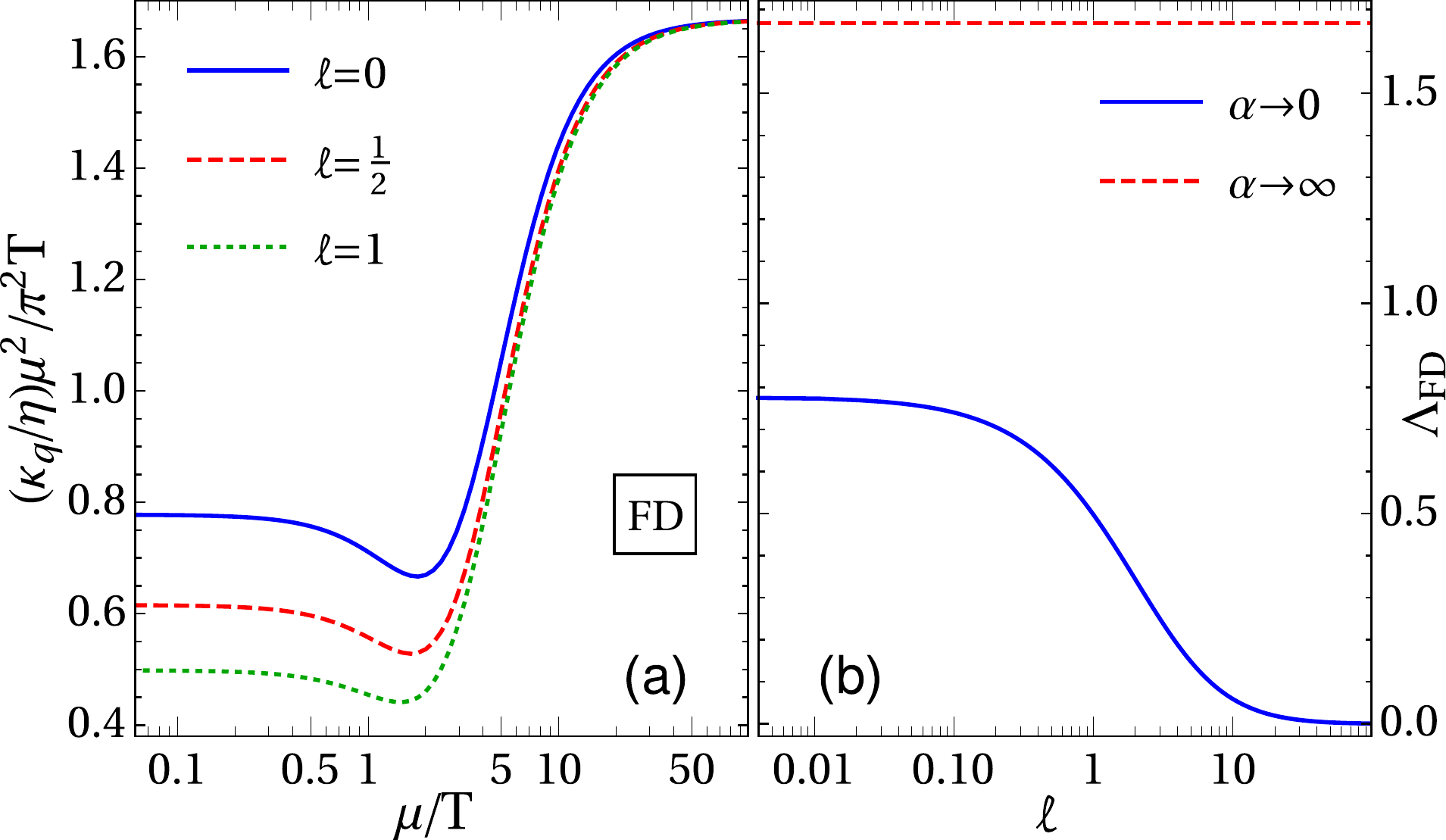}
\end{center}
\vspace*{-6mm}  
\caption{Fermi-Dirac statistics: (a) Evolution of the ratio of heat conductivity to shear viscosity multiplied with the factor $\mu^2/\pi^2T$ as a function of $\mu/T$ for $\ell=0,~1/2$ and $1$. (b) The scaling coefficient $\Lambda_{\rm FD}$ defined in Eq.~\eqref{Lambda_FD} (for $\mu/T\to0$ and $\mu/T\to\infty$) as a function of $\ell$.}
\label{Lambda_FD_alpha}
\vspace*{-.4cm}
\end{figure}

In Fig.~\ref{Lambda_FD_alpha}a, we plot the ratio $(\mu^2\kappa_q)/(\pi^2\eta T)$ as a function of $\mu/T$ for the case of FD statistics for $\ell=0,~1/2$ and $1$. Similar to the MB case, it can be seen that this ratio tends to constant values in the limit of both small and large $\mu/T$, which represents the scaling mentioned in Eq.~\eqref{kappq_eta}. We also observe that all curves with different $\ell$ coincide at large $\mu/T$. This can be easily seen in Fig.~\ref{Lambda_FD_alpha}b where the dashed red line represents $\Lambda_{\rm FD}=5/3$ in the limit $\alpha\to \infty$, indicating that the ratio becomes $\ell$-independent. The solid blue curve representing the evolution of $\Lambda_{\rm FD}$ (in the limit $\alpha\to0$) with $\ell$ shows a monotonous decrease which approaches zero for large values of $\ell$, similar to the behavior observed in Fig.~\ref{Lambda_MB_alpha}b.

\section{Summary and outlook}
\label{sec:summ_conc}

We have derived first-order dissipative hydrodynamic equations using an `extended' Boltzmann equation in the relaxation-time approximation with energy dependent relaxation time. We studied the associated transport coefficients and showed that they acquire corrections due to the energy dependence of relaxation-time. Using a power law parametrization for the energy dependence of the relaxation time, we analyzed ratios of these transport coefficients for different equilibrium statistics and observed several new and interesting scaling features.

Looking forward, it will be interesting to apply the present formulation to derive second-order dissipative hydrodynamics and study the energy dependence of the associated transport coefficients. The framework presented here can also be used to study transport coefficients in other extensions of relaxation-time approximation~\cite{Bhadury:2020ngq}. The present framework can be generalized to derive first-order hydrodynamics equations with different choice of hydrodynamic frames and matching conditions~\cite{Bemfica:2017wps, Kovtun:2019hdm, Bemfica:2019knx, Hoult:2020eho}. Since the current framework allows the relaxation time to be a general function of particle energy, it can be employed to study the scaling of $\zeta/\eta$ with $c_s^2$ for strongly coupled gauge theory plasma \cite{Buchel:2007mf, Buchel:2008uu}. We leave these open problems for future work.

\section*{Acknowledgements}
%
S.J. thanks Chandrodoy Chattopadhyay, Subrata Pal and Jean-Yves Ollitrault, and A.J. thanks Sayantani Bhattacharyya for useful discussions. We thank Alex Buchel for insightful comments on earlier version of this manuscript. S.J. acknowledge National Institute of Science Education and Research, Jatni, for kind hospitality. A.J. was supported in part by the DST-INSPIRE faculty award under Grant No. DST/INSPIRE/04/2017/000038.
%

\bibliographystyle{elsarticle-num}
\bibliography{ref}

\end{document}